\documentclass[twocolumn,aps,prb,superscriptaddress,amssymb,showpacs]{revtex4}
\usepackage[dvips,final]{graphicx}
\begin{document}

\title{Disordered Electrical Potential Observed on the Surface of SiO$_2$ by Electric Field Microscopy} 
\author{N. Garc\'ia}\email{nicolas.garcia@fsp.csic.es}
\affiliation{Laboratorio de F\'isica de Sistemas Peque\~nos y
Nanotecnolog\'ia,
 Consejo Superior de Investigaciones Cient\'ificas, E-28006 Madrid, Spain}
\affiliation{Division of Superconductivity and Magnetism, Institut
f\"ur Experimentelle Physik II, Universit\"{a}t Leipzig,
Linn\'{e}stra{\ss}e 5, D-04103 Leipzig, Germany}
\author{Zang Yan}
\author{A. Ballestar}
\affiliation{Laboratorio de F\'isica de Sistemas Peque\~nos y
Nanotecnolog\'ia,
 Consejo Superior de Investigaciones Cient\'ificas, E-28006 Madrid, Spain}
 \author{J. Barzola-Quiquia}
 \author{F. Bern}
\author{P. Esquinazi}\email{esquin@physik.uni-leipzig.de}
\affiliation{Division of Superconductivity and Magnetism, Institut
f\"ur Experimentelle Physik II, Universit\"{a}t Leipzig,
Linn\'{e}stra{\ss}e 5, D-04103 Leipzig, Germany} 
\begin{abstract}
The electrical potential on the surface of $\sim 300$~nm thick
SiO$_2$ grown on single crystalline Si substrates has been
characterized at ambient conditions using electric field microscopy.
Our results show an inhomogeneous potential distribution with
fluctuations up to $\sim 0.4~$V within regions of $1~\mu$m. The
potential fluctuations observed at the surface of these usual
dielectric holders of graphene sheets should induce strong variations
in the graphene charge densities and provide a simple explanation for
some of the anomalous behaviors of the transport properties of
graphene.
\end{abstract}
 \maketitle

 Nowadays, the study of graphene, i.e. a monolayer of
graphite, represents an important research field in physics and
material science. Although studies of monolayers of graphite grown on
different transitional metal carbides have been already published
nearly 20 years ago \cite{aiz90,ito91}, the simple preparation of
these monolayers by exfoliation \cite{novoscience} as well as grown
on SiC substrates \cite{ber04} and  their transport properties with
the field effect dependence increased substantially the attention of
the solid state community.  One of the highlighted effects is the
electric field-induced metal-insulator-semiconductor transition that
populates the bands of graphene with holes or electrons, bands that
have been claimed to be Dirac like, i.e. following a linear
dispersion relation $E \propto |k|$. Also, the observation of quantum
Hall effects and the quantization of the conductance has been claimed
\cite{gei07}. There are many interesting published results that may
suggest that one can achieve a new electronic basis with this
material if, among other details, one would find a way to produce
homogeneous, uniform layers or with a selective kind of defects
 enabling then their integration in devices \cite{orl08}.

There are, however, some experimental facts indicating that the
transport behavior in graphene is far from being ideal. For example,
the carrier mobility in samples on dielectric substrates including
SiO$_2$ is of the order of 1~m$^2$/Vs, a value that remains rather
independent  of the dielectric substrate, temperature and density of
carriers, see e.g. Ref.~\onlinecite{pon09} and Refs. therein. On the
other hand if the experiments are done with suspended graphene
samples, i.e. without touching the substrate, the mobility
drastically increased \cite{bol08,du08}. The experimental data
suggest that one of the main problems of graphene on dielectric
substrates could come from the substrate non-uniform electrical
potential.

It is generally known that oxides do not exhibit an uniform potential
distribution because a distribution of charges in their near-surface
region exists, letting a metastable charge-potential distribution
 on it. In this case the deposited graphene will be
strongly affected by the same variations of potential the dielectric
substrate has. It is interesting to note the results of an
experimental study using a scanning single electron transistor that
observed puddles of electrons and holes at the graphene surface
\cite{mar08}. The obtained images reveal a rather disordered
domain-like array of fluctuating potential, which might be due to the
substrate influence and not intrinsic of graphene. Further evidence
for the potential influence comes out from the fact that in transport
experiments done in graphene samples one needs to apply a magnetic
field to increase the sample conductivity arguing that otherwise the
carriers are localized \cite{mor08}. In this work we argue that
several of these observations and effects are due to the influence of
the dielectric substrate potential at its near-surface region.
Because of the previous arguments we would like to use the electric
field microscopy (EFM) to analyze the surface of the SiO$_2$ and try
to see if we have potential fluctuations and measure their magnitude.
This will tell us what is the initial state of the potential graphene
sees before shifting the electron and hole bands via a bias voltage.

\begin{figure}[]
\begin{center}
\includegraphics[width=88mm]{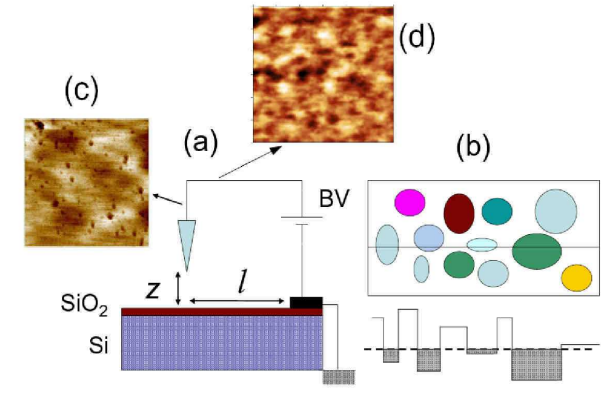}
\caption[]{(a) Sketch of the experimental arrangement. The distance
$z$ between tip and surface can be changed as well as the distance
$l$ to the mass contact. (b) Sketch of the potential distribution to
which a graphene layer would be affected if it is attached to the
surface. The scan line below represents a one dimensional potential
with differently filled wells of graphene carriers. For simplicity
only electron filled bands are depicted. The dashed line represents
the Fermi energy of the graphene layer on top of the disordered
potential surface. (c) EFM picture ($4 \times 4~\mu$m$^2$) of a
SiO$_2$ surface in a sample in which a resin rest (dark spots) were
left. (d) EFM picture ($6 \times 6~\mu$m$^2$) of a
 resin-free sample. These results were obtained with two different microscopes and
different EFM tips. For both EFM pictures the potential gradients
between light and dark broad areas (not spots) are $\lesssim 0.4$~V.}
\label{sketch}
\end{center}
\end{figure}

Consider an EFM arrangement shown schematically in
Fig.~\ref{sketch}(a) where a potential $U_{\rm tip}$ is applied
between the metal tip and the surface of the oxide sample. The
potential difference between tip and surface will be neither zero nor
constant at the surface of SiO$_2$. The applied electric field will
penetrate in the oxide a certain penetration depth $\lambda$ that
depends on the total screening characteristics of the material. Due
to the vanishingly low carrier density of SiO$_2$ it is expected that
the electric field penetration depth $\lambda \propto 1/(n^{1/6}
\sqrt{m^{\star}})$ ($m^\star$ is the effective mass) would be
$\gtrsim 10$~nm for carrier density $n \lesssim 10^{14}~$cm$^{-3}$.
Note that there is a large electrical resistance between the point of
the surface where the tip is and the contact to mass. Therefore the
bias voltage applied $BV$ is between the tip and the thick oxide
layer and the potential drop between the sample tip-position and the
contact on the surface to mass (distance $l$ in
Fig.~\ref{sketch}(a)). Notice however that the last potential drop
will be constant in all the measurements because the scan we perform
is of the order of $5 \times 5\mu$m$^2$ and the distance from the
tip-position to the contact is $l > 0.1$~mm.

The samples we used are usual p-type, polished Si substrates (100)
(Crystec, Germany) with resistivity $\rho \sim 0.02~\Omega$cm) and a
$\simeq 300~$nm thick amorphous SiO$_2$ at the surface grown by
thermal annealing. Some of the substrates were covered by a layer of
insulating optical resin (Pietlow Brandt GmbH, Germany) that was
partially removed with ethanol in some of the samples to investigate
the influence of resin rest on the EFM signal. Other Si substrates
without resin coverage were also measured. The measurements at
ambient conditions were done with two EFM microscopes: an AFM from
NTI Solver and a Dimension 3000 with Extender Electronics Module from
Veeco. The results presented in this work were obtained using the EFM
mode in both. Two different conductive cantilevers were used: Olympus
OMCL-AC240T M-B2, Pt-coated and W2C coated tips with resonance
frequencies around 150.1~kHz and 76.3~kHz. The measurements were
performed in the tapping/lift$^{\rm TM}$ two-pass mode, measuring
first the topography and then the frequency shift of the cantilever
due to the electrostatic force between the tip and surface. At the
first tapping mode no voltage is applied to the tip. If we apply a
constant voltage to the tip and scan the sample surface at a constant
distance from the sample following the track obtained in the first
pass, the measured signal indicates the potential fluctuations on the
sample. In the experiment the frequency shift from resonance depends
linearly on the force gradient given by
\begin{equation}\label{fgs}
\frac{\partial F_z}{\partial z} = \frac{1}{2} \frac{\partial^2
C}{\partial z^2} [U_{\rm tip} - \Psi(x,y)]^2\,,
\end{equation}
where $U_{\rm tip}$ is the voltage difference applied between the tip
of the cantilever and the surface and $\Psi(x,y)$ is the electrical
potential  that interacts with the cantilever tip. It depends on
$\phi(x,y)$ the potential due to charge distribution on the sample
near-surface region and $V_{cp}$ the difference of work functions
between the tip and surface. Within a simple picture one tends to
write $ \Psi(x,y) = V_{cp} + \phi(x,y)$, however both terms of the
r.h.s are interrelated since differences in charge at the surface
would imply also a change in work function. Because the exact value
of $U_{\rm tip}$ is not known due to the further potential drop
within the sample, our results are plotted as a function of the bias
voltage $BV \propto U_{\rm tip}$ that we applied. In the experiments
we obtain an effective force gradient signal in nA units, which is
proportional to force gradient given in Eq.~(\ref{fgs}). As explained
in Ref.~\onlinecite{lu06}, taking the proportionality between these
signals (or their square root) and $BV$, a calibration is done that
is used to transform the measured signals in nA units to voltage
potential changes. In this way any further calibration in terms of
the real force gradient is unnecessary.

\begin{figure}[]
\begin{center}
\includegraphics[width=88mm]{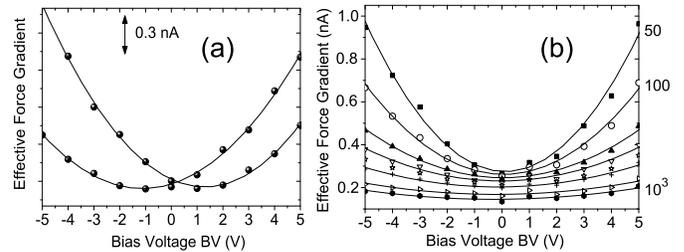}
\caption[]{Effective force gradient in nA vs. Bias Voltage $BV
\propto U_{\rm tip}$. (a) Two curves obtained at a height of 100~nm
between tip and surface at two different positions on the sample. (b)
The same but at different heights in a position of the sample where
the minima is at $BV \simeq 0~$V. The intermediate curves were taken
at heights of 200, 300, 400, 500 and 800~nm. All the continuous lines
are simple quadratic fits to the data in agreement with
Eq.~(\protect\ref{fgs}).} \label{qua}
\end{center}
\end{figure}

Figure~\ref{qua} shows the measured signals vs. $BV$ at different
positions of the SiO$_2$ surface and at different distances $z$. The
result of having the minima of the signal out of zero, see
Fig.~\ref{qua}(a), is not always observed because of the potential
fluctuations. There are other points where the potential falls around
zero as we show in Fig.~\ref{qua}(b). In this case the effective
force gradient signal between -5V and 5V and for tip height between
50~nm and 1000~nm is shown. We observe that all the curves are
centered at $0 \pm 0.15~$V and the curves are practically symmetric
with maximum values for 50~nm and minimum for 1000~nm. The amplitude
of vibration of the cantilever is $\sim 30~$nm and then for $z =
50$~nm the signal depends partially on this vibration. For $z =
1000$~nm the signal is independent of the usual vibration amplitudes
but with a relatively large noise to signal ratio. The best
performances are obtained for 100~nm$ \lesssim z \lesssim 300$~nm.
The results below were obtained for $z \simeq 200~$nm. The results
presented in Fig.~\ref{qua} validate the quadratic dependence given
by Eq.~(\ref{fgs}). The EFM results presented below are therefore
obtained at constant distance $z$ and bias voltage $BV$ and all the
changes in the effective force gradient we measure are due to changes
in the function $\Psi(x,y)$.

\begin{figure}[]
\begin{center}
\includegraphics[width=80mm]{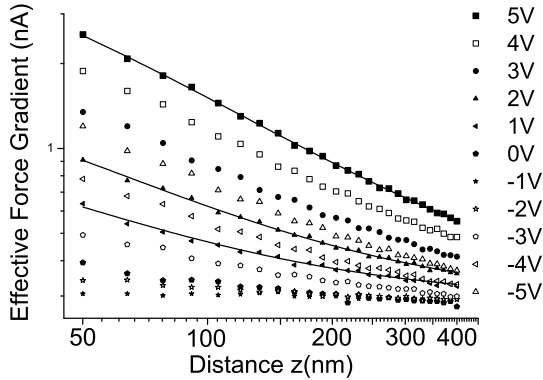}
\caption[]{Effective force gradient in nA vs. distance between tip
and surface in nm taken at different constant applied voltages (see
right list) in a ``bright" region (see e.g. Fig.~\protect\ref{sketch}
or Fig.~\protect\ref{Fig4}). The continuous lines are fit to the
function given by Eq.~(\protect\ref{cond}) with the parameters $A =
380, 100, 56~$nA~nm$^{1.2}$ (for the upper, middle and lower curves),
$\lambda = 25\pm 4~$nm and $b = 1.2$ for the three curves.}
\label{lambda}
\end{center}
\end{figure}

 We have to verify now that the capacitor model
describes the experiments. Consider the prefactor of Eq.~(\ref{fgs}).
If we have a capacitor defined between the tip and surface, then
\begin{equation}\label{cond}
\frac{\partial^2 C}{\partial z^2} = \frac{A}{(z + \lambda)^b} \,,
\end{equation}
where $A$ is a geometrical factor characteristic of the tip-surface
arrangement. In the denominator we have the variation with distance
tip-surface $z$ and $\lambda$ is an effective penetration depth of
the electric field in the SiO$_2$. Note that for a metal $\lambda$ is
practically zero but this is not so for an insulator as shown by the
results presented in Fig.~\ref{lambda} where the EFM signal vs.
distance at constant $BV$ is plotted. If our system has a deviation
from planar electrodes the exponent $b$ in Eq.~(\ref{cond}) should be
smaller than 3. This is what we observe from fitting the results in
Fig.~\ref{lambda} obtaining $b=1.2 \pm 0.04$ and $\lambda = 25 \pm
4~$nm for a region with higher carrier concentration (bright spot)
and $\lambda = 35 \pm 9~$nm for a region with lower carrier
concentration (dark spot), as expected. The dark regions have less
charge and the field should penetrate more.

\begin{figure}[]
\begin{center}
\includegraphics[width=88mm]{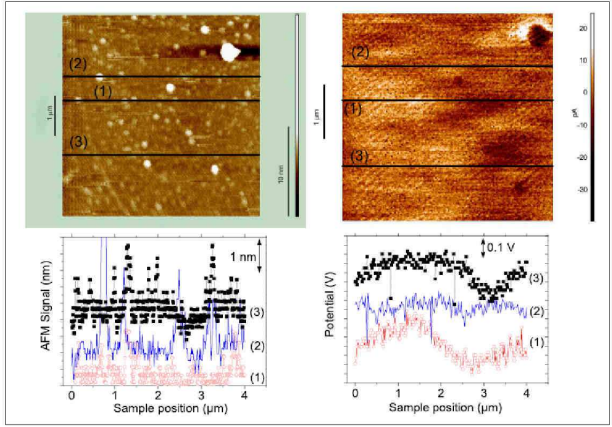}
\caption[]{Left: AFM pictures of the SiO$_2$ surface in a $5 \times
5~\mu$m$^2$ area (the left bar indicates 1~$\mu$m). The pictures
below shows the AFM signal at the three scan lines (1)-(3). Right:
The corresponding EFM result at an area of $3.8 \times 3.8~\mu$m$^2$
inside the area of (a) obtained at $z = 200~$nm and $BV = 3~$V. The
lower picture shows the potential vs. sample position at the three
scan lines. Note that the white spots in the AFM picture correspond
to little resin rests that practically do not produce significant
changes in the EFM signal with exception of the large ones as the one
at the right upper corner.} \label{Fig4}
\end{center}
\end{figure}

We now proceed to take topographical data with AFM and then on the
same scan line EFM data to see the potential variations with respect
to the AFM using the calibration mentioned above. Figure~\ref{Fig4}
shows the AFM (left) and EFM (right) pictures taken on a different
substrate as those shown in Fig.~1(c) and (d) (we have taken more
than 50 scans of this kind in 6~substrates and all look the same).
The white spots correspond to the resin rest that we left to
demonstrate that this rest is not the cause of the relatively broad
variation of the surface potential and it does not prevent of
measuring the topography and potential fluctuations. Similar results
are obtained from a sample without any resin rest, see Fig.~1(d). The
lower pictures in Fig.~\ref{Fig4} show the scans through the lines
(1) to (3) shown in the upper pictures. We observe large fluctuations
in white and black regions through the entire sample. It is clear
from the EFM figures that the oxide surface shows relatively large
potential fluctuations within microns between 0.1~V and 0.3~V for
that sample. Following a similar treatment as in
Ref.~\onlinecite{sch08} this potential can be produced by a charge
density equivalent to $\sim 10^2~$electrons per $\mu$m$^3$. Note that
upon region one can also observe potential fluctuations within a
distance of $0.1-0.2~\mu$m. Note that topographic changes in AFM do
not influence the EFM signal. All the curves obtained between 100~nm
to 600~nm show the same potential variations because these do not
change with the applied voltage indicating that the topography
measurement in the tapping mode scan is reliable.  We checked that
all measured EFM signals provide the complementary contrast by
reversing the voltage polarity applied at the tip indicating that
those signals are due to potential variations and not due to
capacitance artifacts, as explained in detail in
Ref.~\onlinecite{lu06}.

The overall EFM results (Figs.~\ref{sketch} and \ref{Fig4}) mean that
a graphene layer located on these surfaces will feel directly the
potential fluctuations producing regions with higher density of
carriers (electrons as well as holes) at the minima (or maxima) and
extended regions with a smaller carriers' density. In this case the
carriers can be partially localized, see sketch in
Fig.~\ref{sketch}(b), and there will be no conductance unless one
applies a large enough magnetic field or bias voltage as observed
experimentally \cite{mor08}. The influence of these potential
fluctuations on dielectric substrates may also explain the
observation of the quantum Hall effect (QHE), which is apparently not
observed for graphene layers with smaller fluctuations or ``better"
quality. We note also that whereas clear signs of the QHE are
observed in macroscopic HOPG samples \cite{yakovprl03,kempa06} it
appears to be absent in mesoscopic multigraphene samples of good
quality \cite{bartbp}. Another aspect is that the so-called Dirac
point - claimed to have been reached by some experimentalists in
their experiments \cite{gei07} - has not been actually reached. The
carrier density at the Dirac point is several orders of magnitude
smaller that the apparent reached minimum of $\sim
10^{10}$~cm$^{-2}$. Note also that a graphene sample on such a
disordered potential distribution indicates that the use of
 a bias voltage on the transport properties does not assure at all
 a Dirac point crossing through the sample.
 We note that although a linear dispersion
relation for carriers is observed in different spectroscopic
experiments, the existence of a Dirac point at sufficiently low
energies has not yet experimentally proved for graphene/graphite.
Another point is that the number of carriers is difficult to
determine because in real graphene samples and due to the influence
of the substrate and/or attached borders for suspended samples there
will be regions with many and regions without carriers at low enough
temperatures, see Fig.~\ref{sketch}(b).  We would like to remark that
our EFM results have $\sim 0.05~\mu$m resolution. At smaller
distances there could be additional charge distributions that can
also have strong influence on the carrier mobility of the graphene
carriers. The here found inhomogeneous potential distribution on a
dielectric substrate provides a simple way to understand the
experimentally observed constancy of the carrier mobility on
dielectric substrates \cite{pon09}.

In conclusion, EFM measurements on SiO$_2$ surfaces reveal a
disordered potential structure of hills and valleys  similar to those
observed using a single electron transistor on graphene \cite{mar08},
which are due to intrinsic fluctuations of the dielectric substrate.
The potential variations can reach hundreds of mV and therefore the
carriers of graphene attached on a dielectric substrate will have
difficulties to move through the sample affecting their mobility.
These results may explain several unclear behaviors reported in
literature on this topic.

We gratefully acknowledge the  support of the DFG under DFG ES
86/16-1, the DAAD under Grant No. D/07/13369 (``Acciones Integradas
Hispano-Alemanas") and the Spanisch DGICyT. One of us (N.G.) is
supported by the Leibniz Professor fellowship of the University of
Leipzig.


\end{document}